# Machine-learning identification of extragalactic objects in the optical-infrared all-sky surveys

Vladislav Khramtsov,
Department of Astronomy and Space Informatics
V.N. Karazin Kharkiv National University
Kharkiv, Ukraine
vld.khramtsov@gmail.com

Vladimir Akhmetov ,
Laboratory of Astrometry
Institute of Astronomy
V.N. Karazin Kharkiv National University
Kharkiv, Ukraine
akhmetovvs@gmail.com

*Abstract*—We present new fully-automatic classification model to select extragalactic objects within astronomy photometric catalogs. Construction of the our classification model is based on the three important procedures: 1) data representation to create feature space; 2) building hypersurface in feature space to limit range of features (outliers detection); 3) building hyperplane separating extragalactic objects from the galactic ones. We trained our model with 1.7 million objects (1.4 million galaxies and quasars, 0.3 million stars). The application of the model is presented as a photometric catalog of 38 million extragalactic objects, identified in the WISE and Pan-STARRS catalogs cross-matched with each other.

## I. Introduction

Modern space and ground-based astronomical surveys observe ~$10^8$-$10^9$ objects — and amount of observed objects will be increased with time. At the same time, classification of objects, in particular – identification of extragalactic objects, is challenged within such volume of information. Nowadays the main problem is in useless of traditional methods to describe the nature of the order a billion objects in the high-dimensional feature space. Most of the methods of identification extragalactic objects are either defective or non-automatic; such methods use no complete physical or observed information.

Machine-learning algorithms are popular instruments to classify astronomical objects automatically. Here we present new fully-automatic strategy to identify extragalactic objects with analysis of huge amount of sources. Our method has been applied to separate objects into two classes (galactic and extragalactic) with using data from two photometric surveys: space-based mission in mid-infrared Wide-field Infrared Survey Explorer (AllWISE, 700 millions objects; [1,15]) and ground-based telescope in optical and near-infrared Panoramic Survey Telescope and Rapid Response System (Pan-STARRS1, 1.9 billion objects; [2]). The Sloan Digital Sky Survey Data Release 14 catalog (SDSS DR14, [3]) has been used as training sample for our purpose; this catalog contains 4 million spectroscopically confirmed galaxies, quasars and stars.

In result, we got the catalog of extragalactic objects as a realization of our classification method applied to the investigated WISE and Pan-STARRS1 data. We used autoencoder neural network to split-off representative features and Support Vector Machine (SVM) to separate extragalactic objects from stars within constructed representative feature space. Also we used One-Class SVM to provide outliers detection.

We consider our classification method as instrument to quantify amount of galaxies within photometric astronomical catalogs and as independent approach to analyze quality and stellar contamination of catalogs of extragalactic objects.

## II. Classification model

### A. Principles

To make an identification of extragalactic objects it is crucial to create best-suited classification scheme for a given dataset. In our approach, we used SDSS DR14 catalog as a training sample containinig two classes of objects, where each record from this sample is presented in the WISE and Pan-STARRS1 catalogs. With the training sample selected, we have to define a classification scheme working within assembled training sample. Let us consider we have some feature space and we can correctly classify objects into galactic and extragalactic within it; also let we have some function separating objects into two groups. According this, we would define classification model as a set of these two components.

Learning is the main principle of our classification model. We try to tune a set of classifiers to derive three important steps: 1) to build feature space; 2) to detect all anomaly (extramodel) sources; 3) to separate galaxies from stars within built feature space. Also we have to define boundaries divided feature space into two subspaces: model (which limits all possible features of objects from training sample) and extramodel. This assumption is need to avoid classification of objects with unknown (for the constructed model) features.

### B. Data in use

Investigated sample contains objects presented in the two catalogs – WISE and Pan-STARRS. We paired these two catalogs by the spherical coordinates of objects with cross-matching radius equals 1.5 arcsecond; cross-matching of catalogs has been done with using method described in [14].

After pairing and filtration, our investigated sample was contained of 150 million objects. Resulting sample we will call WISExPan-STARRS1.

Further we paired investigated sample with SDSS DR14 and in result got about 2 million objects. Spectroscopic catalog SDSS DR14 contains confirmed stars, galaxies and quasars, (where quasars and galaxies are included in the one class and stars in the another one) so resulting sample is considered to be as a training one.

### III. FEATURE SPACE CONSTRUCTION

Often during identification of extragalactic objects a few parameters only are selected for feature space construction; of cause, these parameters can correctly describe difference between stars and extragalactic objects. But manual selection of features is somewhat arbitrary process and it has to be explained completely. Also manual feature selection tough to produce within huge amount of unknown objects. In addition, the loss of information about behavior of objects within unselected features is promoted. To avoid this, one can use representation learning algorithms.

Representative learning algorithms – it's mostly unsupervised algorithms to analyze latent properties of data. These algorithms have only one aim: they transform input data into special feature space. Further it is possible to describe all properties of objects from the data due to these features. Basic representative learning algorithms are PCA (Principal Component Analysis), LDA (Linear Discriminant Analysis) that use linear data transformation; with using Isomap, LLE (Linear Locally Embedding) one can find complex manifold describing data on the low level with using nonlinear transformation.

In our study we used autoencoder [4] as a representative learning algorithm. Autoencoder is unsupervised method to analyze the latent properties of the data, that can be implemented with neural networks. In the autoencoders representation of data into latent space (encoding) and recovering input data from the latent space (decoding) stages are always realized.

Autoencoder is unsupervised algorithm that means possibility to learn autoencoder to code input data without known classes of objects. It should be clear that autoencoder tries to learn the most informative and independent with each other features. These features can effectively represent input data and can be used as "compressed" input space to classification within it.

Let *x* denote input vector for a training object. During first stage with encoder function *f*, the compressing input vector to the latent vector *h=f(x)* is produced; decoder function *g* recovers input vector from the latent vector as *y=g(h)=g(f(x))*. Autoencoder trains to minimize the loss function *L(x,y)* for the all train batches. The functional form of the loss function have to be chosen according meaning of components of input vectors or due to considered task.

In most of cases, the encoder *f* and decoder *g* functions are neural networks with parameters Θ. Learning of the autoencoder, in neural network terms, rests on finding such Θ provided minimum of the recovering loss.

In our study we used deep autoencoder (Table 1) with 36 input parameters from the training set and mapped it into 5 latent variables (or features). Training set consists of 1.7 millions objects, where we used 1,364,600 samples to learn autoencoder and 341,150 samples to validate learning process. We received mean squared error about 8.0e-05 on the training sample and ≈7.9e-05 on the validation sample after 20 epochs with Adam [9] optimizer. We didn't provide more than 20 epochs to learn autoencoder by dint of components' in the input vector precision ~$10^{-3}$-$10^{-4}$.

TABLE I. AUTOENCODER ARCHITECTURE

| Layer name | Layer description | | |
|---|---|---|---|
| | Neurons | Activation function | Weight initializer |
| Input | 36 | | |
| Encoder_1 | 15 | *tanh* | Orthogonal[a] |
| Encoder_2 | 10 | *elu* | Orthogonal |
| Latent | 5 | *sigmoid* | Orthogonal |
| Decoder_1 | 10 | *linear* | Orthogonal |
| Decoder_2 | 15 | *sigmoid* | Orthogonal |
| Output | 36 | *linear* | Orthogonal |

a. See [5] for more details; *gain = 1.0* and *seed = None* for all initializers

After that we employ derived neural network to predict five encoded features for the WISExPan-STARRS1 sample.

### IV. SUPPORT VECTOR MACHINE SEPARATION

This section describes SVM [6] basics in the view of the our classification model. Main procedure of SVM classification consists of two main steps processed in the feature space: anomaly detection with One-Class SVM and separation of the two classes with kernel-SVM.

#### A. Support Vector Machines

Here we outline the basic idea of the SVM classifier. In the following, we will define that a single feature vector is a *h* = [$h_1, h_2, ..., h_m$]∈ $\mathbb{R}^m$, which is made of *m* latent variables after encoding. The training set {*h*$_1$,$d_1$},{*h*$_2$,$d_2$},…,{*h*$_n$,$d_n$} is composed on *n* objects, where $d_j$ is earthier -1 or +1 indicating the class of *j*-th object.

Let (*w,h*) – *b = 0* denote hyperplane, separating objects of the two classes. Also we have to constrain the position of separating hyperplane in the some stripe between two classes: $d_j$ ((*w,h$_j$*) – *b*) ≥ 1. This can be done only with normalized training data ( $h_i$ is in [-1;1] range for the each *i*).

The optimization problem is to find parameters *w* and *b* corresponding optimal separating hyperplane. This problem can be solved with using classical Lagrangian multipliers method or with using gradient (neural) method. In the first case, we can formulate conditions of optimality as follows: Φ(*w*)= ||*w*||$^2$→ *min* and  $d_j$ ((*w,h$_j$*) – *b*) ≥ 1; solving allow us to

find geometrical meaning of learning – objects for which Lagrange multipliers $α_j≠0$ are support vectors.

Especially important case of non-linear separability. In this deal we can apply the next methods: transforming of original feature space with some positive definite kernel (so-called kernel trick, [7]) or\and giving classifier a chance to make mistakes.

### B. One-Class SVM

We would not to extrapolate result got in SVM separation stage to classify objects with unknown properties, so it is crucial to determine a range of features. Objects with undefined features in the constructed model must to have no class. Determining hypersurface bounding model objects (objects from the training sample) is one of the possible solves of this problem. Anomaly detection in our research has been done using One-Class SVM algorithm [8].

Let we have training sample $H=\{h_1,h_2,…,h_n\}$ which is made of $n$ objects for each of which the feature vector is determining. We want to highlight subsample $S$ consists of $k$ objects so that a previously unseen object $h_{n+1}$ lies in the S with some a priori specified probability $υ$. Parameter $υ$ also determines a maximum fraction of objects from $H$, included in $S$: $υ≥k/n$. The problem may be reformulated as estimating $n$-dimensional function $F$ so that $F$ is positive inside an area limited by subsample $S$ (and respectively is negative elsewhere).

This problem can be reformulated in the SVM terms. Let us consider the training sample $H$ as a set of objects of one class and the origin as coordinates of object from the second class. The task can be reduced from the one-class problem to the binary one – in result, we want to build separation hyperplane between objects from $H$ and origin. This formulation has no practical sense without kernel trick implementation. Since this reformulation we can solve one-class problem with standard binary SVM realization, adding only one free parameter $υ$.

### C. Results

We applied OCSVM to the train data (1.7 million sources) with Radial Basis Function kernel: $K(x_i,x_j)=\exp[-γ\|x_i-x_j\|^2]$ (RBF) with a free parameter $γ$. We used n-fold cross-validation method to determine optimal $γ$ and $υ$ within the next ranges: $log(υ)$ between -5 and 0, $log(γ)$ between -5 and 2. We found that the best parameters are $υ=10^{-5}$ and $γ=1.0$, for which negligible fraction of objects from training sample (23 objects only) is expected to be anomaly. OCSVM classifier applied to the WISExPanSTARRS1 data returns only 0.5 million objects that are expected to be anomaly (Fig.1).

Further we applied SVM classifier to divide training sample into galactic and extragalactic objects. Our training sample consists of about 300,000 stars, so we randomly choose 300,000 extragalactic object to make balanced sample. We wouldn't to use weights within all (1.4 galaxies and 0.3 million stars) unbalanced sample because we don't know a real observed fraction of objects of these classes. As in previous step, we used n-fold cross-validation technique to determine free parameters of classifier (with RBF kernel transformation) $C$ and $γ$. We determined the best score of separation for the next parameters' ranges: $log(C)$ between -2 and 2; $log(γ)$ between -5 and 2. In result, we got that C=10.0 and $γ$=100.0 provide the least separation error (Table 2) equals about 0.47%.

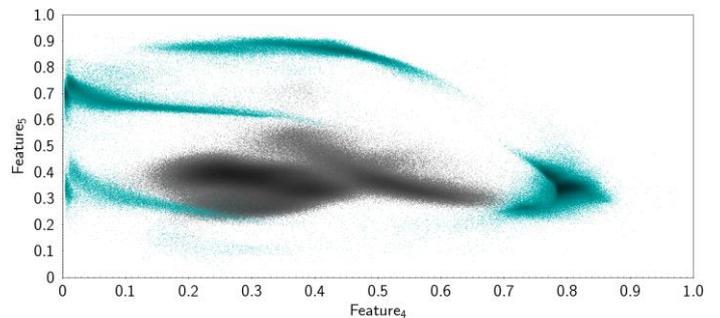

Fig. 1. Distribution of objects from train sample (1.7 million objects, gray dots) and anomalies, detected within WISExPan-STARRS1 with OCSVM (0.5 million objects, blue dots) on the (Feature$_4$-Feature$_5$) plane

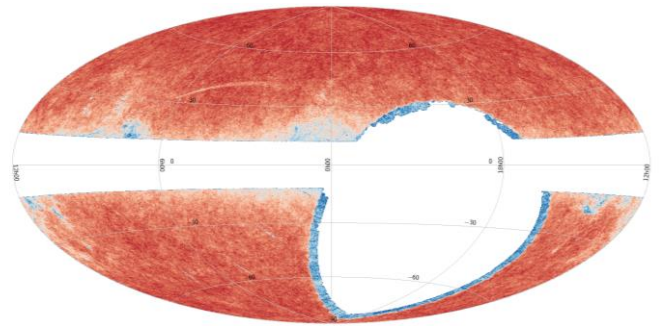

Fig. 2. Distribution of sources classified as extragalactic in the WISExPan-STARRS sample

TABLE II.  SVM RESULTS

| Metric | Extragalactic objects | Galactic objects |
|---|---|---|
| Precision | 0.99284 | 0.99865 |
| Completeness | 0.99866 | 0.99279 |
| F-measure | 0.99574 | 0.99572 |

After constructing separating hyperplane in the feature space, and also hypersurface bounding model features, we applied our classification model to the whole WISExPan-STARRS1 data. In result, we found that 38 million objects from the 150 million investigated ones are extragalactic (Fig.2).

## V. CONCLUSIONS

In this work we demonstrated application of automated feature selection, anomaly detection and classification in task of star-galaxy separation. By using the our classification model, trained and tested on a cross-match of spectroscopic SDSS DR14 data with WISExPan-STARRS1, we identified about 38 million extragalactic objects.

Our classification model is automatic approach to analyze catalogs of extragalactic objects and can be applied to identify extragalactic objects in any set by dint of learning.

Successful machine-learning identification of extragalactic objects within WISExPan-STARRS1 shows that constructed classification model can be applied to the other sky surveys.

Catalog of identified extragalactic objects available at http://astrodata.univer.kharkov.ua/astrometry/db/.

ACKNOWLEDGMENT

This publication makes use of data products from the Wide-field Infrared Survey Explorer, which is a joint project of the University of California, Los Angeles, and the Jet Propulsion Laboratory/California Institute of Technology, funded by the National Aeronautics and Space Administration.

The Pan-STARRS1 Surveys have been made possible through contributions by the Institute for Astronomy, the University of Hawaii, the Pan-STARRS Project Office, the Max-Planck Society and its participating institutes, the Max Planck Institute for Astronomy, Heidelberg and the Max Planck Institute for Extraterrestrial Physics, Garching, The Johns Hopkins University, Durham University, the University of Edinburgh, the Queen's University Belfast, the Harvard-Smithsonian Center for Astrophysics, the Las Cumbres Observatory Global Telescope Network Incorporated, the National Central University of Taiwan, the Space Telescope Science Institute, and the National Aeronautics and Space Administration under Grant No. NNX08AR22G issued through the Planetary Science Division of the NASA Science Mission Directorate, the National Science Foundation Grant No. AST-1238877, the University of Maryland, Eotvos Lorand University (ELTE), and the Los Alamos National Laboratory.

Funding for the Sloan Digital Sky Survey IV has been provided by the Alfred P. Sloan Foundation, the U.S. Department of Energy Office of Science, and the Participating Institutions. SDSS acknowledges support and resources from the Center for High-Performance Computing at the University of Utah. The SDSS web site is www.sdss.org.

SDSS is managed by the Astrophysical Research Consortium for the Participating Institutions of the SDSS Collaboration including the Brazilian Participation Group, the Carnegie Institution for Science, Carnegie Mellon University, the Chilean Participation Group, the French Participation Group, Harvard-Smithsonian Center for Astrophysics, Instituto de Astrofísica de Canarias, The Johns Hopkins University, Kavli Institute for the Physics and Mathematics of the Universe (IPMU) / University of Tokyo, Lawrence Berkeley National Laboratory, Leibniz Institut für Astrophysik Potsdam (AIP), Max-Planck-Institut für Astronomie (MPIA Heidelberg), Max-Planck-Institut für Astrophysik (MPA Garching), Max-Planck-Institut für Extraterrestrische Physik (MPE), National Astronomical Observatories of China, New Mexico State University, New York University, University of Notre Dame, Observatório Nacional / MCTI, The Ohio State University, Pennsylvania State University, Shanghai Astronomical Observatory, United Kingdom Participation Group, Universidad Nacional Autónoma de México, University of Arizona, University of Colorado Boulder, University of Oxford, University of Portsmouth, University of Utah, University of Virginia, University of Washington, University of Wisconsin, Vanderbilt University, and Yale University.

*Software*: LIBSVM [10], Tensorflow [11], TOPCAT [12], scikit-learn [13]